# Demonstrating backflow in classical two beams' interference

Anat Daniel[†], Bohnishikha Ghosh[†], Bernard Gorzkowski[†], Radek Lapkiewicz

*Institute of Experimental Physics, Faculty of Physics, University of Warsaw, Ludwika Pasteura 5, 02-093 Warsaw, Poland*

b.ghosh@uw.edu.pl, anat.daniel@gmail.com , bernard.gorzkowski@gmail.com , radek.lapkiewicz@fuw.edu.pl
[†]*These authors contributed equally to this work*

**The well-known interference pattern of bright and dark fringes was first observed for light beams back in 1801 by Thomas Young. The maximum visibility fringes occur when the irradiance of the two beams is equal, and as the ratio of the beam intensities deviates from unity, fringe visibility decreases. An interesting outcome that might not be entirely intuitive, however, is that the wavefront of such unequal amplitude beams' superposition will exhibit a wavy behavior. In this work, we experimentally observe the backflow phenomenon within this wavy wavefront. Backflow appears in both optics (retro-propagating light) and in quantum mechanics, where a local phase gradient is not present within the spectrum of the system. It has become an interesting subject for applications as it is closely related to superoscillations whose features are used in super resolution imaging and in a particle's path manipulations. The first successful attempt to observe backflow was made only recently in an assembly of optical fields, by synthesizing their wavefront in a complex manner. Yet, backflow is perceived as hard to detect. Here, by utilizing interference in its most basic form, we reveal that backflow in optical fields is robust and surprisingly common, more than it was previously thought to be.**

The phenomenon of backflow was discovered more than 50 years ago by quantum physicist G.R Allcock[1] while studying the problem of arrival times in quantum mechanics (QM). In his theoretical work, he uncovered a counterintuitive effect that occurs when a forward propagating particle, with only positive momenta, has a non-zero probability of propagating backwards during specific time intervals. Bracken and Melloy, following his study, provided an upper limit to the probability of backflow, giving a broader perspective on the phenomenon[2]. Despite the strange nature of the effect or perhaps due to it, backflow has appeared in a volume of research spanning over myriad disciplines such as, Bose-Einstein Condensate (BEC)[3], black holes[4], quantum mechanical particles[5,6,7,8,9], etc. The common understanding in all of these works is that backflow is, just as in the double slit interference of matter waves, a manifestation of the wave nature of quantum particles. Other theoretical studies[10,11] show that backflow is a versatile phenomenon that occurs in any system supporting coherent waves interference, including quantum mechanical wave-functions (where the probability density flows backwards) and classical waves (where the energy density flows backwards). Here, we shed light on the classical wave-like picture of backflow by introducing a simple experimental demonstration.

Effectively, backflow is present when the local Fourier components of the superposition are not contained within its Fourier spectrum[5], an outcome often referred to as superoscillation[10,12,13,14]. In the latest experimental demonstration of backflow in optical systems, Eliezer et al[12], designed a beam of light which manifests backflow in the transverse momentum by creating a complex superposition of plane waves with only negative Fourier components using a spatial light modulator (SLM). The authors found positive local Fourier components by scanning a slit in the Fourier plane of the SLM. In their work, the observation of backflow requires precise control over the input state of the beam. In our work, on the other hand, backflow arises naturally from an optical interference of two beams with unequal amplitudes, in free space. The first advantage of this approach is that the maximum amount of backflow is no longer limited by the spatial resolution of the SLM. The second advantage is its generic nature: in order to observe backflow, one does not need to tailor a complex wavefront but only superpose two wavepackets—a requirement fulfilled in various physical systems such as classical electromagnetic waves, neutrons[15], molecules[16], electrons[17], and BEC[18].

We initially show analytically how backflow arises from the wavy wavefront that is a result of the superposition of two unequal beams. Fig. 1 depicts the intensity distribution and the wavefront (yellow curves) of superposition of two plane waves with unequal amplitudes, equally inclined to the z-axis $\psi(x,z) = e^{i(z+ax)} + be^{i(z-ax)}$, where $b$ is the ratio between the amplitudes, $2a$ is a measure of the angle between the propagation direction of the plane waves, x and z are the transverse and longitudinal directions respectively (all parameters are dimensionless). The wavy nature of the wavefront[19,20] (Fig. 1(a)) leads to strong phase gradients $\vec{k} = \vec{\nabla}\arg\{\psi(x,z)\}$ in the dark fringes, and thus, a higher x-component of the local wave vector (red arrows, orthogonal to the wavefronts) than those of the component plane waves in these regions (white arrows). The dark fringes are hence the regions in which backflow can be detected (see for example, region marked by the green box). In the case of perfectly equal amplitudes however, backflow does not occur (Fig. 1(b)). We note that the longitudinal i.e., z components of the constituent plane waves are assumed to be equal. This assumption could be violated, and would lead to the observation of backflow in the longitudinal direction. However, in our demonstration, we focus on backflow in the transverse direction only.

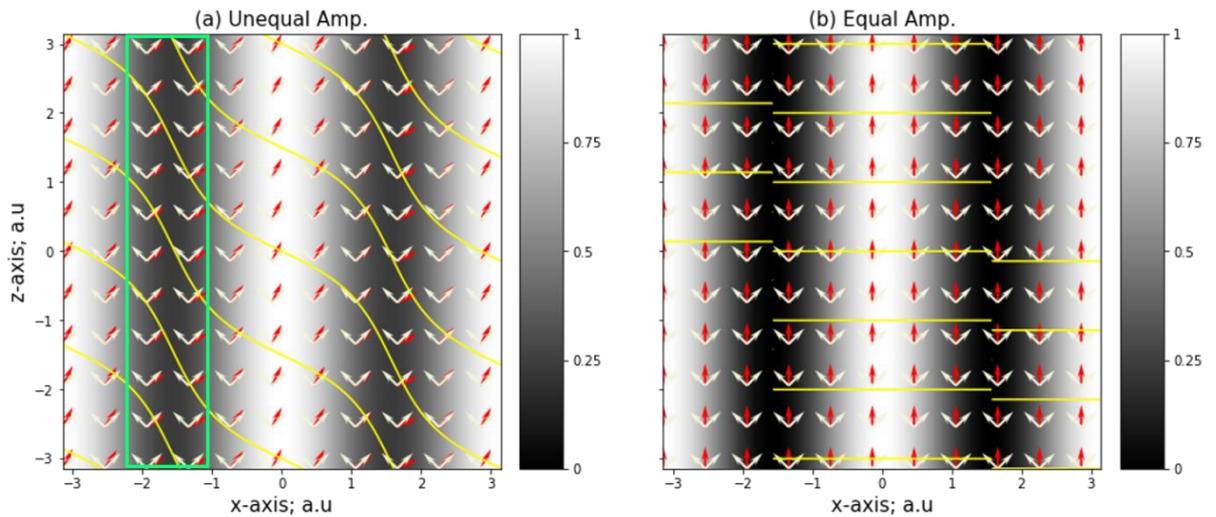

**Fig. 1: Visualization of backflow in unequal two plane waves interference, along the propagation direction (simulation).** The intensity distribution of the superposition of two plane waves is given by $|\psi(x,z)|^2 = 1 + b^2 + 2b\sin(2ax)$ (gray scale map), where b is the ratio between the amplitudes, 2arctan(a) is the angle between the propagation direction of the plane waves. The yellow curves represent the wavefronts given by $\arg\{\psi(x,z)\} = $ const. We have chosen $a = 1$. The local wave vector

of the superposition (arrows in red) is given by $\vec{k}_s = \vec{\nabla} \arg\{\psi(x,z)\} = \left(\frac{a(1-b^2)}{1+b^2+2b\sin(2ax)}, 1\right)$. The wave vectors of the constituents [$\vec{k}_1 = (a,1), \vec{k}_2 = (-a,1)$] are plotted with white arrows. The arrows are normalized i.e., they have the same length. (a) Unequal amplitudes, $b = 0.35$. The wavy behavior of the wavefront is evident. In the region of the dark fringes, the red arrow $\vec{k}_s$ lies outside the "triangle" formed by $\vec{k}_1$ and $\vec{k}_2$, as its $x$-component is higher than those of the constituents, corresponding to backflow (see an example of backflow region marked by the green box). (b) Equal amplitudes, $b \to 1$. The wavefronts are flat with phase jumps due to singularities in the zero intensity lines of the fringes. In contrast to (a), here $\vec{k}_s$ lies in between $\vec{k}_1$ and $\vec{k}_2$, in all regions of space. No backflow is seen. Note that while defining backflow here, we do not distinguish between the positive and negative directions of the x-axis but merely consider the exceeding of x-component of the wave vector from those of the constituents.

Aforementioned, backflow refers to the counterintuitive flow of energy density or probability density and thus, the accepted description of backflow would be using the *Poynting vector*[11,21]. Nevertheless, given that we consider scalar fields in free space, the Poynting vector $P(\mathbf{r}) = \text{Im}\{\psi^*(\mathbf{r})\nabla\psi(\mathbf{r})\} = |\psi(\mathbf{r})|^2 \nabla \arg\{\psi(\mathbf{r})\}$, is merely the local wave vector scaled by intensity[19,20], thereby making the demonstration in wave vector-space, which we adopt here, an indication of the direction of energy flow.

Plane waves are infinite in extent and are therefore experimentally infeasible. Hence, we prepare the superposition of two Gaussian beams which are made to interfere at their waists (Fig. 2 (a)), wherein their wavefronts are planar. Additionally, the region of interference is magnified, thus emulating a superposition of plane waves, up to a small but finite width of the spectrum of each beam (for further details on backflow in Gaussian beams see Supplementary Material B).

Fig. 2 (b) illustrates the experiment setup. A superposition of two Gaussian beams is prepared by using a polarization based Mach-Zehnder interferometer. A 780 nm polarized laser beam passes through a polarizing beam displacer (PBD, ThorLabs BD40) that separates its orthogonal polarizations. In order to control the amplitude ratio between the beams (i.e controlling $b$ in $\psi(x,z)$), a half wave plate is placed prior to the PBD1. A second polarizing beam displacer (PBD2), together with a half wave plate rotated by 45 degrees in the basis of PBD1, compensates for the path difference of the two beams and allows us to control the transverse separation between them. The beam separation determines the angle of intersection (i.e controlling $a$ in $\psi(x,z)$). Vertical fringes are ensured for every beam separation by rotating the optical axis of PBD1 w.r.t. the z axis. A polarizer is set to 45 degrees in the basis of PBD2. A lens (L1, *f*=150 mm) combines the beams at its focal plane and a cross-section of the interference pattern is imaged onto a Shack-Hartmann wavefront sensor, consisting of a microlens array (ThorLabs-MLA-150-5C-M) attached to the CMOS camera (mvBlueFOX-200wG). The initial beam waist of 0.14 mm is magnified by a factor: M=28.6, to make the beams' spectrum narrower, such that the CMOS sensor of dimension 4 mm intercepts approximately the waist. The local wave vectors of the superposition are measured by the Shack-Hartmann wavefront sensor technique[22].

By virtue of vertical fringes on the camera, we ensure that there is only one transverse component of the wave vector. Yet, in general, in case of two transverse components, the predicted backflow remains unchanged (for further details see Supplementary Material A).

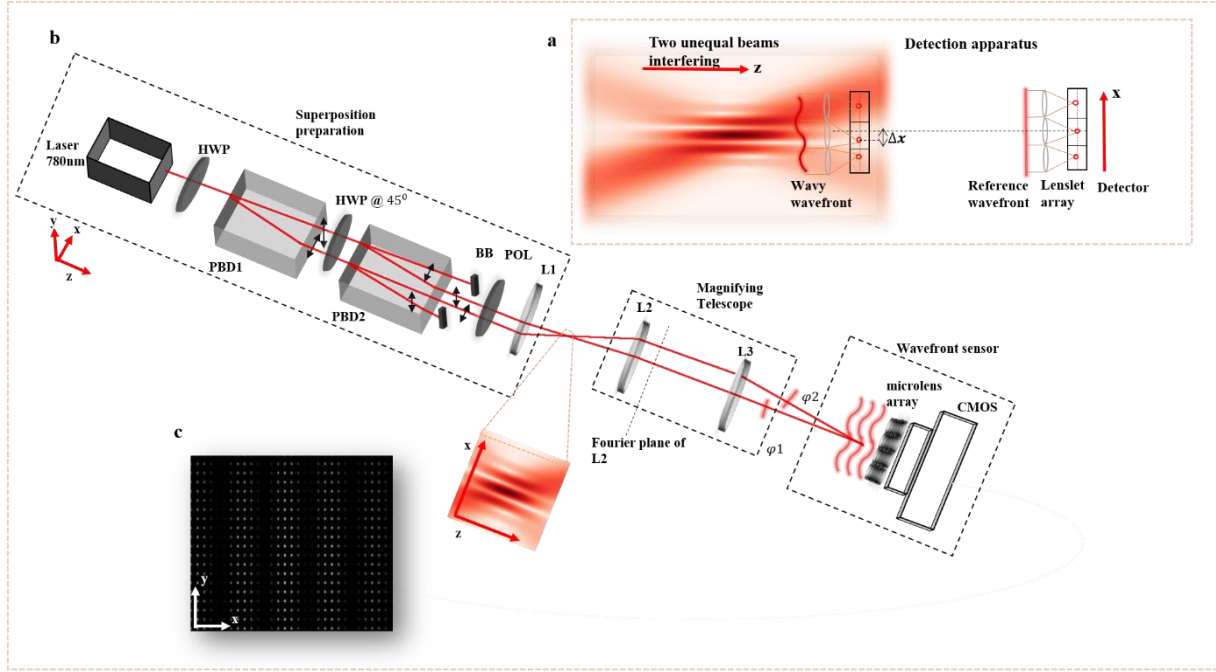

**Fig. 2: Experiment setup for demonstrating backflow.** (a) Concept of the setup. Two magnified Gaussian beams of unequal amplitudes intersect at their waists (simulated). Owing to unequal intensities of the beams, the wavefront of the resultant superposition is wavy (not flat). Following the Shack-Hartmann technique, by finding the centroids of the spot-field and measuring their displacement $\Delta x$ with respect to a reference, the wavefront of the beams' superposition is reconstructed. The local wave vector is derived for each lenslet. (b) A polarization-based Mach-Zehnder interferometer for superposing two beams. Half wave plate (HWP) sets the ratio between the two beams. A polarized beam displacer (PBD1; unrotated w.r.t. the lab frame) separates the initial beam into its horizontal and vertical components. A second polarizing beam displacer (PBD2) together with a half wave plate rotated by 45 degrees compensate for the path difference of the two beams and set them with a fixed transverse separation. Two additional beams that emerge from PBD2 are blocked (BB). A polarizer (POL) is inserted to enable the interference of the two orthogonally polarized parallel beams that emerge from PBD2, followed by a lens (L1, $f$=150 mm) that combines them. The inset is a simulation the interference pattern of two beams along the axis of propagation−z. A *4f* imaging system (L2, $f$=35 mm and L3, $f$=1000 mm), with a magnification of 28.6, images the a xy cross section of the interference pattern on to the wavefront sensor. $\varphi_1$ and $\varphi_2$ indicate the wavefronts of beam 1 and beam 2 respectively, the beams' wavefronts and their resultant superposition wavy wavefront are measured by the wavefront sensor. The spectrum of beam 1 and 2 is measured at the focal plane of lens L2 by placing the CMOS camera without the microlens array at this plane. (c) The spot field of the xy cross section of the interference fringes, given by the microlens array.

As expected from the wavy wavefront, the plot in Fig. [3] shows the oscillating wave vector of the superposition (magenta curves). Backflow is observed in the transverse $x$ direction, in regions where the local wave vector of the superposition $k_{s,x}$ exceeds its constituent Gaussian momentum distributions, i.e., the spectrum of beam 1 (centered at $k_{1,x}$, width $3\sigma$) and of beam 2 (centered at $k_{2,x}$, width $3\sigma$). This result can be associated with Fig. [1], where $k_s$ exceeds the triangle formed by the mean directions of beam 1, $k_1$, and beam 2, $k_2$, thus leading to backflow. We estimate $k_{s,x}$, $k_{1,x}$ and $k_{2,x}$, by finding the centroids of the spot-field[23] (Fig. [2] (c)) and measuring their displacement with respect to a reference. The reference in our setting is not measured directly, but calculated: corresponding to each lenslet, the centroid position of the spot given by each beam is obtained, and their mean position is assigned as the reference position. The diffraction spread induced by each lenslet does not change the position of the centroid thereby assuring that it is determined only by the local momentum.

The value of *a* is extracted by fitting the intensity cross section to a sinusoidal function with a Gaussian envelope (details in the caption Fig. [3]). The amplitude ratio between the beams, *b,* is obtained by fitting the Gaussian function to the images of the constituent beams and taking the ratio of the Gaussian amplitudes. These parameters are then used to calculate the theoretical prediction of $k_{s,x}$. The experimentally measured $k_{s,x}$ is in excellent agreement with the theoretical prediction. It may be observed that the heights of the peaks corresponding to backflow are slightly different across the x position, this is owed to a slight shift between the centers of the constituent Gaussian beams; see Supplementary Material C where we derived the theoretical $k_{s,x}$ including the aforementioned correction. Next, we obtain the spectrum of the superposition by taking its Fourier transform (Fig. 3 inset), using a lens (lens L2 in Fig. 2). We note that due to measuring the spectrum (local transverse momentum) by implementing the Fourier transform with a lens (lenslet), the demonstration is confined to the paraxial approximation[24].

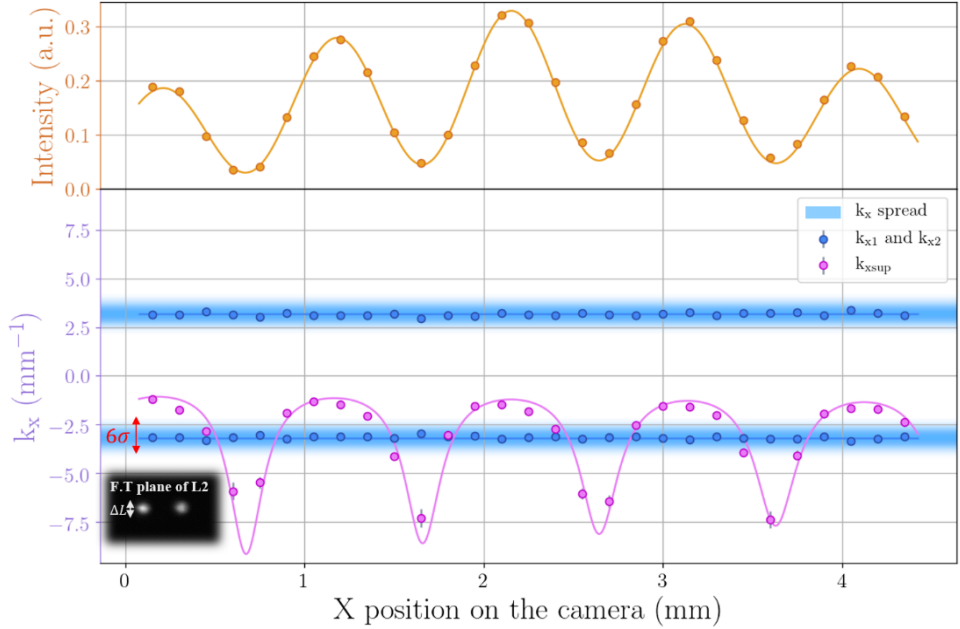

**Fig. 3: Experimental observation of backflow.** The x-axis is the x position across the lenslet-array (mm). The orange dots represent experimental values of intensity across each lenslet. The orange curve is a fit of the experimental data to a sinusoidal function with a Gaussian envelope $e^{-\frac{2x^2}{w_0^2}}(1 + b^2 + 2b\cos(2ax))$ where $w_0$ is the width of the Gaussian beams, $b$=0.45 is the ratio measured in the experiment. From this fit $a$ is extracted $a$=3.19 mm$^{-1}$. The blue and the magenta dots represent the experimental values of $k_{1,x}$, $k_{2,x}$ and $k_{s,x}$ (mm$^{-1}$) respectively, extracted from one row of the spot field matrix (as seen in Fig. 2(c)). The error of each data point in a given x position is found by estimating the standard deviation of centroids in the corresponding column of the spot field matrix. The blue lines, $k_{1,x}$ and $k_{2,x}$ fit the data to a constant value, as expected from the behavior of Gaussian beams at their waists. The magenta curve represents the theoretical prediction of $k_{s,x}$ (see caption of Fig. 1 and Supplementary Material C). The wave vector components are found by measuring centroid displacement $\Delta x$ with respect to the reference, and scaling appropriately: $k_x = \frac{2\pi\Delta x}{\lambda f_m}$; $\lambda$ is the wavelength, and $f_m$ is the focal length of each microlens. The inset shows the Fourier transform of the two beams for estimating the width $\pm 3\sigma$ of their spectrum by fitting a Gaussian function. The blue shaded region is $k_{1,x} \pm 3\sigma'$. As the lens used for the Fourier transform ($f_{FT}$) and the lenslets have different focal lengths, $\sigma = \frac{2\pi\Delta L}{\lambda f_{FT}}$ is appropriately scaled to $\sigma' = \frac{\sigma}{M}$. , where $M$ is the magnification of the imaging system. In the dark fringes, the experimental values of $k_{s,x}$, while being in good agreement with the theory, are seen to exceed not only $k_{1,x}$ (the x-component of the wave vector of the brighter beam) but also the spread of the Fourier transform.

In order to confirm that we observed backflow, it is required to estimate the contribution of the momentum values which lie in the tail of the Gaussian spectrum of the individual beams, and show that it is negligible compared to the "anomalous" local momentum values arising from the superposition. For this estimation, the backflow probability $P_{BF}$ is given by integrating the intensity over the regions of x positions within one fringe where anomalous momentum values are observed, and dividing by the intensity over one period of the fringe. The probability of finding the anomalous values within the Gaussian tails of the spectrum $P_{SP}$ is given by integrating the spectrum over the part of its tail that can lead to such anomalous momentum values, and dividing by the integral over the full spectrum[5] (see [Supplementary Material D](#) for the exact calculation). For the data presented in Fig. [3](#) we find that $P_{BF}$ is bigger than $P_{SP}$ by 3 orders of magnitude, hence confirming backflow.

The spectrum may also obscure and limit the observation of the backflow, however this is not a fundamental limitation as we can further magnify the Gaussian beams thereby narrowing their spatial spectra.

As a further insight, we experimentally study the dependence of backflow on the amplitude ratio (*b*) and the angle between the two superposed beams (*a*). Fig. [4](#) shows the features of backflow in a given interval of x. Fig. [4](#) (a-c) is a comparison between cases wherein the ratio *b* is constant but the angle between the beams is altered. As expected from an increased angle, the number of dark fringes per unit distance increases. Hence, the appearance of backflow is more frequent but the spatial extent of each peak is smaller. That is to say, for large angles the x-component of the local wave vector has a better chance to exceed the spectrum (i.e. backflow), but its detection requires finer sampling in the x direction. Fig. [4](#) (d-f) is a comparison between cases wherein the angle between the beams is constant but the ratio *b* is altered. Here, regardless of the ratio, the backflow appears with the same frequency across the interval of *x*. However, when *b* is closer to unity (but *b*≠1; otherwise, we reach a singularity in the dark fringe) there is still a better chance that the backflow exceeds the spectrum as the peaks are higher; simultaneously they get narrower and might not be sampled owing to the spatial resolution of the measuring device. We repeat the procedure of confirming backflow for the data in Fig. [4](#) (a-f) and find that in all the cases except (d) $P_{BF}$ is bigger than $P_{SP}$ by 3 orders of magnitude; while in (d) we do not confirm backflow, as seen clearly from the plot.

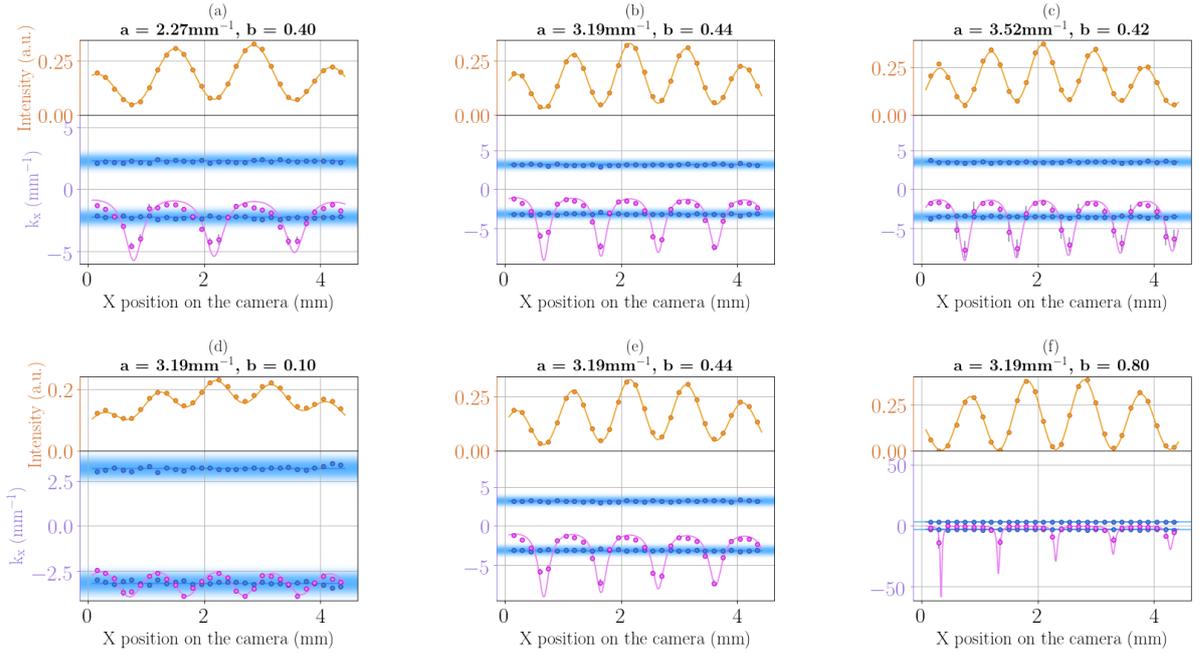

**Fig. 4: The effect of the amplitude ratio and the angle between the beams on the observation of backflow.** (a-c) Ratio, *b*, is constant and the angle between the beams is increasing. As the angle increases, the displacement due to the backflow is higher and its appearance is more frequent. Namely, the observed backflow has a better chance to exceed the spectrum for higher angles, but simultaneously, its detection requires finer sampling in the x direction and therefore higher spatial resolution. (d-f) The angle between the beams, *a,* is constant and the amplitude ratio is increasing. Note how the interference contrast changes. As *a* is the same in all the cases, backflow appears with the same frequency across the interval of *x*. On the other hand, the displacement due to backflow is greater as the ratio between the beams increases. When *b* is closer to unity (but *b*≠1*;* otherwise, we reach a singularity in the dark fringe) the x-component of the local wave vector is more likely to exceed the spectrum (i.e., backflow) as the peaks are higher, but might not be sampled owing to the spatial resolution of the measuring device. A case where backflow is not observed is (d), where the displacements do not exceed the spectrum. In order to observe it, one can magnify further the Gaussian beams and accordingly shrink their spectra. In (f) the changing heights of the theoretically predicted peaks are due to the non-overlapping beams as mentioned in the main text and in the Supplementary Material C. Yet, the observation of backflow is not affected.

In summary, we demonstrate backflow by measuring the local transverse component of the wave vector of the superposition of two wide Gaussian beams, and show that, in the dark intensity regions, it exceeds the spectrum of its constituents. The interference takes place in free space thereby excluding effects related to anisotropy or dispersion of media. Our study on optical waves makes use of a simple experimental configuration and can possibly be extended to various types of systems including matter waves[15,16,17,18], single photons[25,26], and mechanical waves. Apart from preparing a superposition of two wavepackets, observation of backflow requires a measurement of local momentum, which is relatively straightforward and involves only spatial filtering and detection in the far field. In the case of single photons, for example, we could repeat the experiment and expect the results to be consistent with the current ones achieved using classical beams.

Backflow thus far was considered to be exotic and difficult to observe. Here, on the contrary, we show that it would be hard not to observe backflow as it is experimentally infeasible to satisfy the criterion of equal intensity of the interfering beams. Additionally, we can control the parameters (i.e., amplitude ratio and angle) and understand their physical relevance in observing the effect. Unlike the previous study[12] which involves constraints on state preparation (for example the resolution of the SLM), we show that it is not necessary to engineer a state that manifests backflow a priori, as such an engineering might be difficult in particular physical systems. The two-dimensional single shot local momentum measurement using a Shack-Hartmann wavefront sensor, devoid of scanning, can be advantageous in systems manifesting backflow in any two transverse directions, for example in beams containing orbital angular momentum or any azimuthal degree of freedom[20,27,28,29,30]. Moreover, our setup enables studying backflow for partially coherent superpositions (corresponding to mixed quantum states) as local transverse momenta of partially coherent light can be measured with a Shack-Hartmann wavefront sensor[31]. However, this awaits further theoretical investigations.

As a transition from being merely a fundamental phenomenon under study, to also having possible applications, backflow can be found to be useful, for instance, in tailoring electromagnetic beams with non-conservative forces[32,33], and in sub-diffraction limit imaging technologies[34,35]. Once the effect is accessible and easy to achieve in the lab, as we have demonstrated here, it opens pathways for other potential applications involving interference of classical and quantum mechanical wavepackets.

## Acknowledgements

We thank Arseni Goussev and Iwo Bialynicki-Birula for valuable insights and discussions that broadened our understanding of the field. We are indebted to Yaron Silberberg for inspiring the research on wavy wavefronts and Tomasz Paterek for introducing us to the concept of backflow and for insightful comments on our manuscript. This work was supported by the Foundation for Polish Science under the FIRST TEAM project 'Spatiotemporal photon correlation measurements for quantum metrology and super-resolution microscopy' co-financed by the European Union under the European Regional Development Fund (POIR.04.04.00-00-3004/17-00).

## Supplementary Material

### A. <u>Theoretical description of optical backflow in the interference of two plane waves with equal longitudinal components</u>

Note that in the main manuscript we consider the superposition of two plane wave components with equal but opposite x-components and the same z-component of the wave vector. We have thus far neglected the y-component of wave vectors of these plane waves. The following is the superposition of two plane waves with equal z-components of the wave vector. However, the x and y components can be in general different.

$$\Psi(x,y,z) = e^{iz}\left(e^{i(a_1 x + b_1 y)} + b e^{i(a_2 x + b_2 y)}\right) \qquad (1)$$

Note that here $a_1^2 + b_1^2 + 1 = a_2^2 + b_2^2 + 1$ and that $a_1, b_1, x, y, z$ are dimensionless here.

A manipulation of equation (1) leads to the following expression.

$$\Psi(x,y,z) = e^{iz}e^{i\frac{(a_1+a_2)x}{2}}e^{i\frac{(b_1+b_2)y}{2}}\left(e^{i[(a_1-a_2)x+(b_1-b_2)y]/2} + be^{-\frac{i[(a_1-a_2)x+(b_1-b_2)y]}{2}}\right) \quad (2)$$

Let us define $\frac{a_1-a_2}{2} = a'; \frac{b_1-b_2}{2} = b'$

Therefore, we can simplify equation (2) in the following manner.

$$\Psi(x,y,z) = e^{iz}e^{i\frac{(a_1+a_2)x}{2}}e^{i\frac{(b_1+b_2)y}{2}}\left(e^{i\sqrt{a'^2+b'^2}\left[\frac{a'}{\sqrt{a'^2+b'^2}}x+\frac{b'}{\sqrt{a'^2+b'^2}}y\right]} + \right.$$
$$\left. be^{-i\sqrt{a'^2+b'^2}\left[\frac{a'}{\sqrt{a'^2+b'^2}}x+\frac{b'}{\sqrt{a'^2+b'^2}}y\right]}\right) \quad (3)$$

Let us now define a rotation of coordinates in the x-y plane such that $\frac{a'}{\sqrt{a'^2+b'^2}}x + \frac{b'}{\sqrt{a'^2+b'^2}}y = x'$, where $x'$ is the new x-axis.

Hence, equation (3) simplifies to the following expression.

$$\Psi(x,y,z) = e^{iz}e^{i\frac{(a_1+a_2)x}{2}}e^{i\frac{(b_1+b_2)y}{2}}\left(e^{i\sqrt{a'^2+b'^2}x'} + be^{-i\sqrt{a'^2+b'^2}x'}\right) \quad (4)$$

In equation (4), we observe that barring the phase offsets, the superposition consists of plane wave components with equal but opposite inclinations to the z-axis in the $x - z$ plane. Hence, instead of the generic structure of equation (1), it suffices to consider the following simpler scenario.

$$\Psi(x,z) = e^{iz}(e^{iax} + be^{-iax}) \quad (5)$$

Since the y component of the wave vector can be neglected up to a rotation, it is not significant. Nonetheless, we ensure that the y component of the wave vector is zero by making the optical elements parallel to the table and at incidence at the same height. Vertical fringes ensure this.

In the following section, we consider equation (5) with a Gaussian amplitude envelope.

### B. Theoretical description of optical backflow in the interference of two Gaussian beams

The superposition of Gaussian beams with an amplitude ratio of $b$ is given as

$$\Psi_S = (e^{iax} + be^{-iax})\frac{w_0}{w(z)}e^{-\frac{r^2}{w^2(z)}}e^{i\left(\frac{kr^2}{2R(z)}-\phi(z)+kz\right)} \quad (6)$$

Where, $w(z) = w_0\sqrt{(1 + (z/z_R)^2)}$, is the waist of the beam as a function of the position along the direction of propagation, $\frac{1}{R(z)} = \frac{z}{(z^2+z_R^2)}$ represents the inverse of the curvature and $z_R = \frac{kw^2_0}{2}$ is the Rayleigh range of the beam and depends on beam parameters such as wave-number and the beam waist at $z = 0$. The Rayleigh range is a measure of how fast the curvature of the beam changes near its waist. The $\phi(z) = \arctan\left(\frac{z}{z_R}\right)$ is the Gouy phase that appears from the propagation along z. Note that for the sake of simplicity, all the quantities involved in equation (6) are dimensionless.

A linear phase in x between the beams is chosen in order to reduce the problem to one dimension and thereby focus on the 'backflow' in this direction.

The local wave vector is found by taking the gradient of the phase of the state. $\vec{k} = \vec{\nabla} \arg\{\Psi_S\}$.

It is evident from the symmetry with respect to the z axis that the z component of the wave vector ($k_z$) will remain unchanged. It is the x component of the local wave vector that plays a role here. Focusing on z=0 allows to cancel out the curvature dependent terms in the expression of the local wave vectors of the superposition and the individual beams. Table 1 lists the expressions of the x-components ($k_x$) of the local wave vectors of the superposition and the individual beams. The x-component of the local wave vector of the superposition contains two terms, the first of which depends on the curvature of the beam and can be owed to its Gaussian nature. The second term can be owed to the factor in brackets in equation (1) and is due to the gradient of the phase of the superposition of two plane waves equally but oppositely inclined to the z-axis. The denominator is the intensity for the superposition of plane waves. The intensity distribution of the Gaussian beams' superposition should have a Gaussian envelope. The fact that this term increases as we increase the angle between the beams ($a$) or make the ratio $b$ closer to unity, is of interest to us. We have tried to verify the dependence of these parameters experimentally.

|       | $\vec{\nabla}\arg\{\Psi_s\}$ | $\vec{\nabla}\arg\{\Psi_1\}$ | $\vec{\nabla}\arg\{\Psi_2\}$ |
|---|---|---|---|
| $k_x$ | $\frac{kx}{R(z)} + \frac{a(1-b^2)}{1+b^2+2b\cos(2ax)}$ | $\frac{kx}{R(z)} - a$ | $\frac{kx}{R(z)} + a$ |
| $k_z$ | $\partial_z \left(\frac{kr^2}{2R(z)} - \phi(z) + kz\right)$ | | |

**Table 1**: The expressions of local wave vector components for each Gaussian beam component and their superposition, as described by equation (1). The leftmost column represents x and z components of the local wave vector. The z component is insignificant as it is equal for each beam and the superposition. The x component of the wave vector of the superposition can be greater than the x component of the wave vector of the beam with higher intensity in the region of the dark fringes. For the sake of simplicity, we assume that we look at z = 0 as the curvature dependent term $\frac{kx}{R(z)}$ disappears in this region. The individual beams would thus have constant $k_x$s in this region.

It must be noted that we focus on the configuration of the beams that are equally inclined to the z axis in opposite directions (disregarding the y direction) for the sake of the simplicity that this configuration offers.

While we do not distinguish between the positive and negative directions of the x-axis but merely consider the exceeding of x-component of the wave vector from the spectrum, the original concept of optical retro propagation/ optical backflow as described by MV Berry[10] makes use of beams traveling to the right (positive x) and local momentum being found on the left (negative x), analogous to the original idea of quantum backflow.

### C. Theoretical description of optical backflow in the interference of two Gaussian beams with their centers shifted by 2ξ

Let us consider the superposition of two plane waves of unequal amplitudes with Gaussian envelopes with their centers shifted in equal and opposite directions along $x$ by $\xi$.

$$\Psi_S(x,y,z) = e^{iax}\frac{w_0}{w(z)}e^{-[(x+\xi)^2+y^2]/w^2(z)}e^{i[k\{(x+\xi)^2+y^2\}/2R(z)-\phi(z)+kz]} + be^{-iax}\frac{w_0}{w(z)}e^{-[(x-\xi)^2+y^2]/w^2(z)}e^{i[k\{(x-\xi)^2+y^2\}/2R(z)-\phi(z)+kz]}, \quad (7)$$

where $w_0, k, w(z), R(z), \phi(z)$ are as defined in Supplementary section B. Here $a$ and $x$ are in the units of inverse of length and length respectively. When the Gaussian envelopes have a large Rayleigh range and plane of interest is $z=0$, equation (7) can be approximately modified to the following.

$$\Psi_S(x,y,z) \approx e^{ikz}e^{-\frac{x^2+y^2+\xi^2}{w_0^2}}\left[e^{-\frac{2x\xi}{w_0^2}}e^{iax} + be^{\frac{2x\xi}{w_0^2}}e^{-iax}\right] \quad (8)$$

Thus, the phase and its x gradient can be expressed as follows.

$$\arg\{\Psi_S(x,y,z)\} = z + \arctan\left[\frac{1-B(x)}{1+B(x)}\tan(ax)\right], \quad (9)$$

where $B(x) = be^{\frac{4x\xi}{w_0^2}}$ is the position dependent ratio between the beams.

$$k_x = a\frac{e^{-\frac{4x\xi}{w_0^2}} - b^2 e^{\frac{4x\xi}{w_0^2}} - 4b\frac{\xi}{aw_0^2}\sin(2ax)}{e^{-\frac{4x\xi}{w_0^2}} + b^2 e^{\frac{4x\xi}{w_0^2}} + 2b\cos(2ax)} \quad (10)$$

The factor $\frac{b4\xi}{aw_0^2}$ in the experiment is approximately 0.02b for all values of $a$. Therefore, the effect of the non-overlapping beams can be considered to be negligible for ratios that are not close to unity. However, for b close to unity, the effect on the heights of the peaks is visible (as can be seen in Fig. 4 (f)); nonetheless, backflow still persists and can be detected with a high confidence.

In Fig. 3 and 4 of the main text the heights of the peaks of the backflow change across the x position due to local variations in the amplitude ratio between the beams ($B(x)$). Although, the degree of brightness changes across the camera, for ratios much less than unity, one beam is consistently brighter than the other, and hence the peaks have the same signs but are gradually smaller in heights.

However, when the ratio is closer to unity (b=0.98) one beam is brighter than the other across only half of the camera and hence the signs of the peaks flip in addition to being different in heights. See Fig. S1.

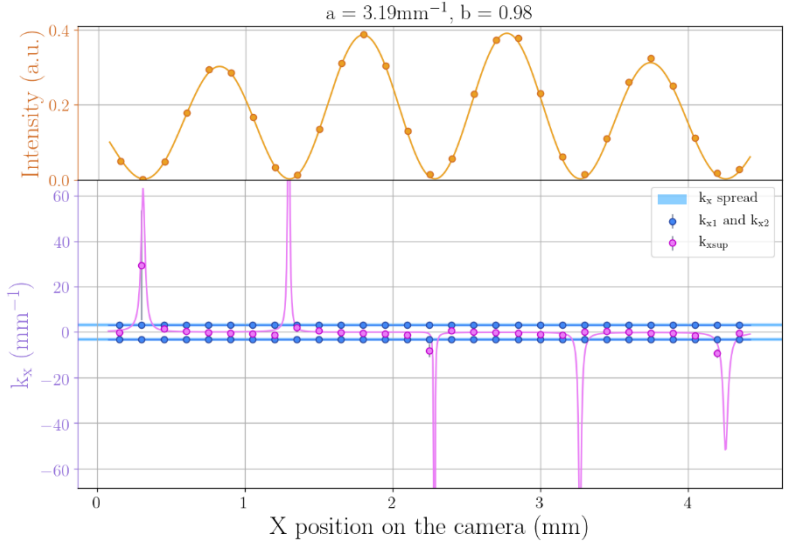

**Fig. S1: The effect of non-overlapping beams on the sign of the peaks of backflow when the amplitude ratio is very close to unity.**

### D. Estimating the probability of backflow $P_{BF}$ and the probability of finding the anomalous values within the Gaussian tails of the spectrum $P_{SP}$

The probability of backflow is given by:

$$P_{BF} = \frac{\int_{x_0}^{x_0 + \text{fringe size}} \int_{-\infty}^{\infty} I(x,y) \mathbb{1}_{k_x < k_{x,ref}}(x) dy dx}{\int_{x_0}^{x_0 + \text{fringe size}} \int_{-\infty}^{\infty} I(x,y) dy dx} \quad (11)$$

The probability of finding the anomalous values within the Gaussian tails of the spectrum is given by

$$P_{SP} = \frac{\int_{-\infty}^{k_{x,ref}} \int_{-\infty}^{\infty} e^{-\frac{\lambda^2 f^2}{4\pi^2 w_0^2}\left((k_x - k_{x0})^2 + k_y^2\right)} dk_y dk_x}{\int_{-\infty}^{\infty} \int_{-\infty}^{\infty} e^{-\frac{\lambda^2 f^2}{4\pi^2 w_0^2}\left((k_x - k_{x0})^2 + k_y^2\right)} dk_y dk_x} \quad (12)$$

where $I(x,y)$ is the intensity distribution, $x_0$ is the position of an arbitrary fringe, $k_{x,ref}$ is the point we choose to cut the tail of the gaussian, $k_{x0}$ is the wave vector of one of the initial beams, $f$ is the focal length of the lens used for the Fourier imaging, $w_0$ is the waist of the gaussian in the Fourier plane, $\mathbb{1}_{k_x < k_{x,ref}}(x)$ is the identity function (equals 1 when $k_x$ is lower than $k_{x,ref}$, 0 elsewhere).